\documentclass[useAMS,usenatbib]{mn2e}
\usepackage{psfig}
\newcommand{\velasnr}{{RX J0852.0-4622}}

\newcommand{\SII}{{[S{\sc\,ii}]~6716~\&~6731~\AA}}
\newcommand{\oiii}{[O~{\sc iii}]~5007\AA}
\begin{document}
\title[RCW 37 and \velasnr]{Kinematics of the Pencil Nebula (RCW 37) and 
its association with the young Vela supernova remnant \velasnr}
\author[M.P. Redman, J. Meaburn, M. Bryce, D.J. Harman and T.J. O'Brien]
{M.P. Redman$^{1}$, J. Meaburn$^{2}$, M. Bryce$^{2}$, D.J. Harman$^{2}$, and T.J. O'Brien$^{2}$\\
$^1$ Department of Physics \& Astronomy, University College London,
Gower Street, London WC1E 6BT, UK.\\ $^2$ Jodrell Bank Observatory,
University of Manchester, Macclesfield SK11 9DL, UK.}
\date{\today}
\pubyear{2001} 
\volume{000}
\pagerange{\pageref{firstpage}--\pageref{lastpage}}
\maketitle 
\label{firstpage}

\begin{abstract}
The association between the Pencil nebula (RCW 37, NGC 2736), the Vela
X-ray fragment D/D' and the recently discovered new X-ray supernova
remnant (\velasnr) in Vela is investigated. Recently published Chandra
images of D/D' are compared with optical images of RCW 37 and confirm
the close association of the two objects. New optical line profiles of
RCW 37 from an extended slit position passing through this unusual
optical nebula are presented. They reveal a partial velocity ellipse
with expansion velocities of around $120~{\rm km~s^{-1}}$. Various
scenarios for the origin of the nebula are considered and the evidence
of a link with \velasnr\ is reviewed. A funnel of gas similar to those
in the Crab and DEM34a SNRs is not ruled out but a more plausible
explanation may be that a `wavy sheet' is responsible. We suggest that
\velasnr\ is located within the older larger Vela SNR
and that some of the X-ray gas from \velasnr\ has collided with the
dense H~{\sc i} wall of the older remnant. This gives rise to the
morphology and velocity structure of the optical emission and explains
the unusual X-ray emission from this portion of the supernova
remnant. If our hypothesis is correct, a distance prediction of
$250\pm 30~{\rm pc}$ can be made, based on recent measurements of the
distance to the old Vela SNR. This is at the lower end of the range of
distances quoted in the literature would confirm unusual nature of
this young nearby supernova remnant.
\end{abstract}


\section{Introduction}
\velasnr\ is a young nearby supernova remnant (SNR) recently discovered 
\citep{aschenbach98,iyudin.et.al98} near the southeastern perimeter of 
the well known old Vela SNR. \velasnr\ has generated much interest since
the distance and age could be as low as $200~{\rm pc}$ and $700~{\rm
yr}$ respectively and thus it could have been generated by the nearest
supernova explosion in recent human history. 

\begin{figure}
\psfig{file=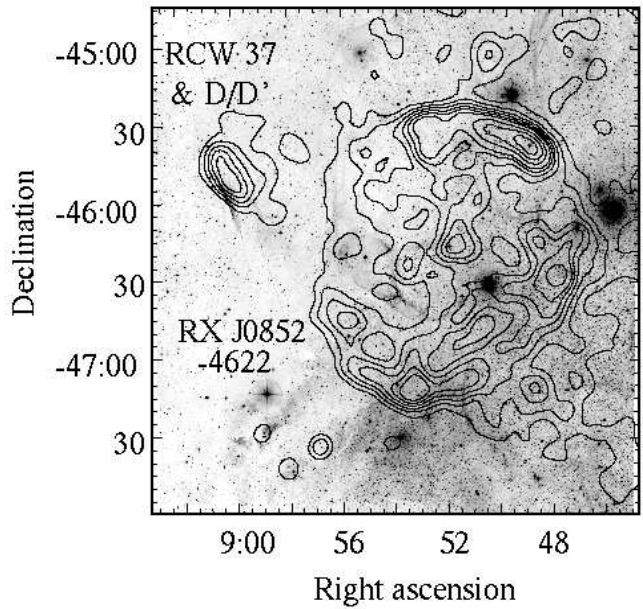,width=250pt,bbllx=78pt,bblly=213pt,bburx=517pt,bbury=629pt}
\caption{Contour map of \velasnr\ from the RASS hard X-ray discovery data 
of \citet{aschenbach98} overlaid on ESO IIIaJ optical images. The
Pencil nebula RCW 37 and X-ray fragment D/D' coincide to the upper
left of the picture.}
\label{esorass}
\end{figure}

The SNR was discovered in Rosat hard X-ray data (shown as contours in
Fig.\@~\ref{esorass}) and at these energies has a shell-like
morphology. There is no obvious optical counterpart to the main body
of the SNR but \citet{redman.et.al00} showed that a fragment of X-ray
emission (labelled D/D' by \citealt{aschenbach.et.al95}) coincides
closely with a bright optical nebula, RCW 37. The X-ray fragment is
located just beyond the main circular body of the remnant and is
clearly visible in hard X-ray images. The main X-ray shell is not
complete and there is a break in the emission in a direction
coincident with that of the X-ray fragment and with RCW 37 (see
Fig.\@~\ref{esorass}). \citet{redman.et.al00} suggested that RCW 37 is
physically associated with \velasnr\ and represents a venting of hot
gas from the interior of the remnant to beyond the roughly circular
shell as delimited in the X-ray.

\begin{figure}
\psfig{file=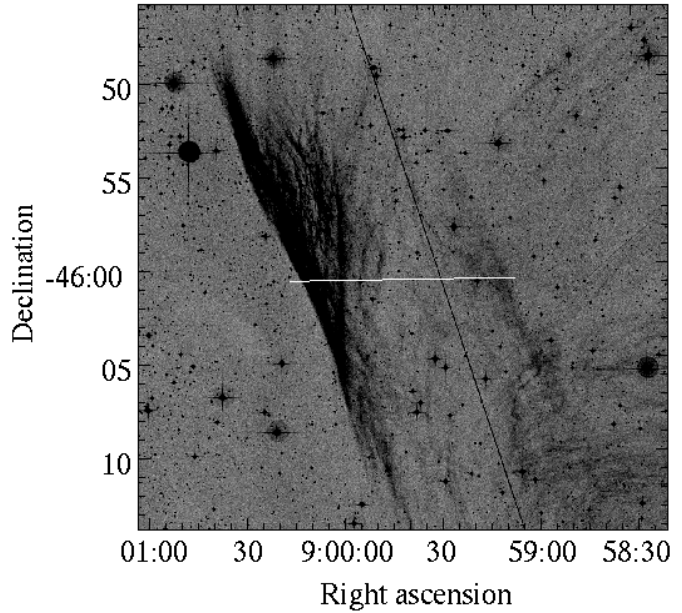,width=250pt,bbllx=64pt,bblly=201pt,bburx=530pt,bbury=641pt}
\caption{ESO image of RCW 37 with the five overlapping MES slit positions
marked with a white line. The sloping thin dark line
is a satellite trail.}
\label{eso}
\end{figure}

RCW 37 (NGC 2736) was discovered in the 1840s by Sir John Herschel and
is a bright optical nebula known to amateur astronomers as the Pencil
Nebula (Fig.\@~\ref{eso}). The unusual, intricate morphology of the
nebula and its large size and brightness make it surprising that there
have been few studies of this object. Its apparent location at the
dust-obscured eastern side of the six degree diameter old Vela SNR, away
from the photogenic western filaments may be one
reason. \citet{blair.et.al95} used the Ultraviolet Spectrometer on the
Voyager 2 spacecraft to investigate O~{\sc vi} emission from RCW 37
and inferred that shock speeds of $160-300~{\rm km~s^{-1}}$ are
present. This is consistent with the limited line profiles obtained by
\citet{redman.et.al00} at the edge of the nebula. The morphology of
the nebula is striking and led \citet{redman.et.al00} to suggest that
its funnel-like appearance could be taken to indicate that there is
indeed a collimated flow of hot gas taking place from the remnant
interior. Alternatively, a pre-existing cloud could be being shocked
by escaping hot gas. The kinematics were not sufficient to establish
clearly the dynamics of the nebula.

In this paper, new optical forbidden line profiles from RCW 37 from an
extended slit position across the filamentary bulk of RCW 37 are
presented in order to determine the structure and dynamics of the
nebula. The kinematics are discussed and used to describe a model for
the origin of RCW 37 that places it in the context of \velasnr\ and
the main Vela SNR. It is argued that \velasnr\ is embedded within the
older larger Vela SNR.

\begin{figure}
\psfig{file=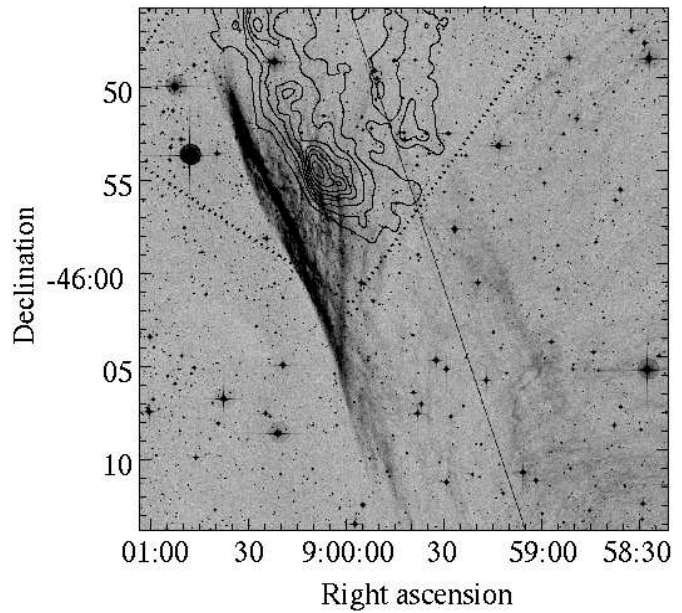,width=250pt,bbllx=65pt,bblly=202pt,bburx=531pt,bbury=640pt}
\caption{ESO image of RCW 37 with contours of the Chandra ACIS X-ray data of 
\citet{plucinsky.et.al02} overlaid. The thick dotted line indicates the edge of
the Chandra data. The sloping thin dark line is a satellite trail.}
\label{esochandra}
\end{figure}

\section{Observations and Results}

Contours of the smoothed hard ($E>1.3~{\rm keV}$) RASS discovery data
of \velasnr\ are shown in Fig.\@~\ref{esorass} overlain on an ESO
archive IIIaJ (green) image of the region. Fig.\@~\ref{eso} is an ESO
archive green image of RCW 37. Contours of smoothed Chandra ACIS data
of D/D' obtained by \citet{plucinsky.et.al02} are shown in
Fig.\@~\ref{esochandra} overlain on an ESO archive green image of RCW
37.

Spatially resolved, longslit echelle spectra of the \oiii\ and \SII\
emission lines were obtained with the Manchester echelle spectrometer
(MES; \citealt{meaburn.et.al84}) combined with the f/7.9 Cassegrain
focus of the Anglo--Australian telescope (AAT). The observations were
made on 2001 February 13-14 during reasonable seeing. A 90~\AA\ wide
interference filter was used to isolate the echelle order containing
the emission lines. A Tektronix CCD with 1024$\times$1024 24~$\mu$m
square pixels was the detector. A slit width of 150 $\mu$m ($\equiv
11~{\rm km~s^{-1}}$ and 1\arcsec) was used.  The CCD was binned by a
factor of two in the spatial direction to give 512 $\times$ 0\arcsec
.32 pixels along the slit length.  Each integration was debiased and
wavelength calibrated to better than $0.5~{\rm km~s^{-1}}$ using
Thorium--Argon spectra obtained between exposures.  Data were obtained
from five slightly overlapping east-west slit lengths to produce an
effective slit length of about 700\arcsec. The slit position is marked
on Fig.\@~\ref{eso}. The data from each slit was reduced and the five
were mosaiced together.

\subsection{Kinematics}
\begin{figure*}
\psfig{file=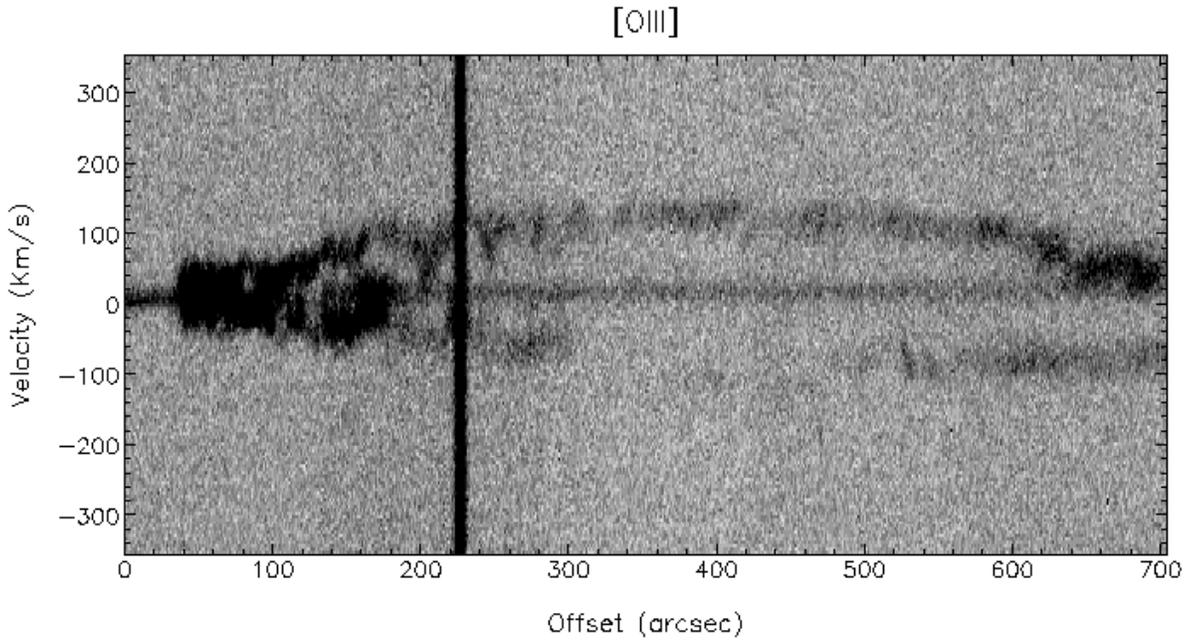,width=450pt,bbllx=37pt,bblly=280pt,bburx=558pt,bbury=562pt}
\caption{Position-velocity array of \oiii\ line profiles mosaiced from five 
consecutive slit positions. The dark vertical line is the spectrum of a star 
intersected by the slit. The vertical axis is heliocentric radial velocity} 
\label{oiiifig1}
\end{figure*}

\begin{figure*}
\psfig{file=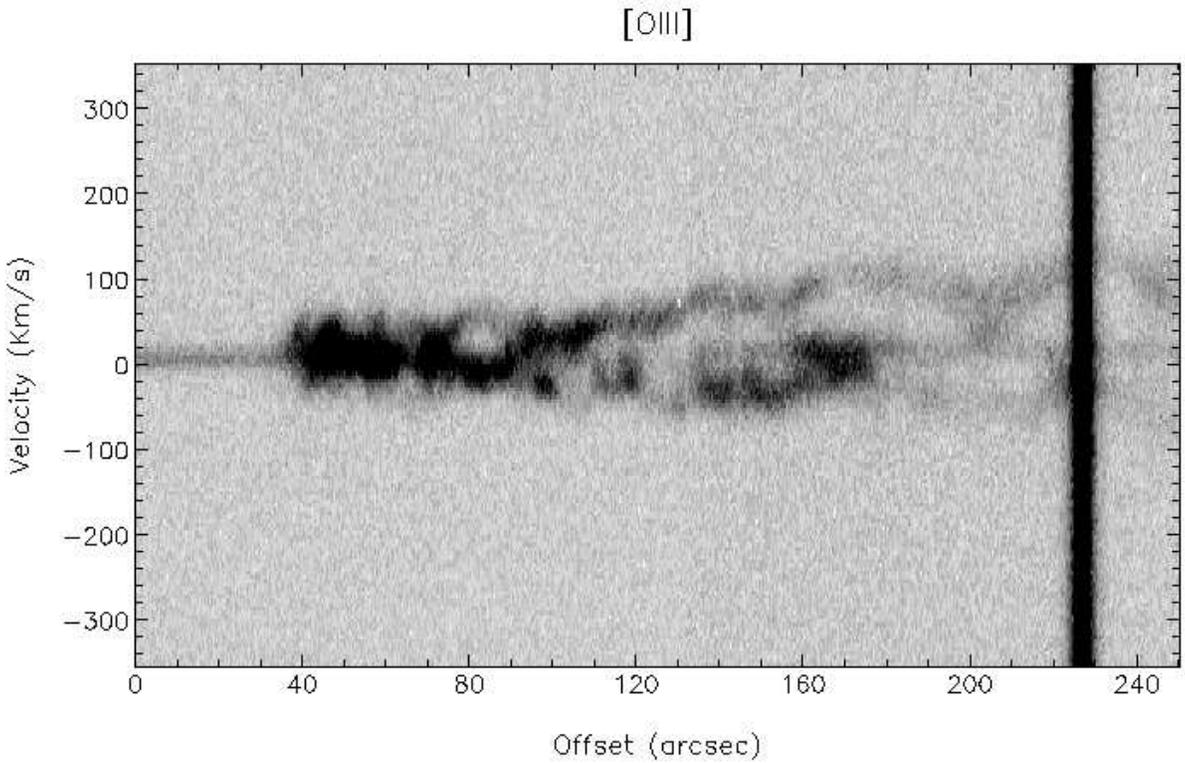,width=450pt,bbllx=38pt,bblly=252pt,bburx=558pt,bbury=590pt}
\caption{Data from previous figure, cropped and rescaled to show the brightest
velocity features at the eastern edge of RCW 37. The dark vertical
line from the intersected star can be used as a reference point
between the two figures. The vertical axis is heliocentric radial
velocity}
\label{oiiifig2}
\end{figure*}

A negative gray-scale representation of a position-velocity (PV) array
of \oiii\ line profiles is displayed in Figs~\ref{oiiifig1} and
~\ref{oiiifig2}. Data from the full effective slit length is displayed
in Fig.\@~\ref{oiiifig1} while Fig.\@~\ref{oiiifig2} is the section of
the data from the eastern side of the effective slit. The bright
vertical band is the spectrum of a star intersected by the slit which
can be used as a reference point between the two figures. The
following features can readily be seen. Throughout the data at zero
heliocentric radial velocity there is faint background \oiii\
emission. The bright eastern edge of the nebula begins at an offset of
around 40\arcsec. The data here exhibit velocity knots and small
velocity loops that correspond to the complex filaments readily seen
in Fig.\@~\ref{eso}. With increasing offsets, the line profiles begin
to split. At offsets of greater than about 300\arcsec\ the profiles are
no longer split and only a single faint component at positive
velocities (up to $\sim 120~{\rm km~s^{-1}}$) is present. The positive
velocity component then approaches the systemic with increasing offset
and we identify the bright feature at around 700\arcsec\ as being the
western edge seen in Fig.\@~\ref{eso}. There is a negative velocity
component from offsets of 550\arcsec\ onwards that we attribute to the
irregular diffuse emission present towards the western side of the
nebula and not associated with RCW 37. 
 

\section{Discussion}

\subsection{Dynamics of the nebula}
The position-velocity arrays show a coherent velocity structure across
the nebula. An almost complete classical velocity ellipse is clearly
apparent. The morphology of the nebula is reminiscent of an incomplete
funnel (or tube) but could also be simply a curved sheet of
emission. This is discussed below but, broadly speaking, the velocity
data indicate gas that is expanding radially at about $100~{\rm
km~s^{-1}}$. At the edge of the nebula, the motion is into the plane
of the sky while towards the middle there is a velocity splitting
between the near and far side of the emitting gas that increases
towards the middle of the nebula. Eventually, the red-shifted emission
disappears while that the blue-shifted emission returns to the
systemic. At this point there is background emission (seen in
Fig.\@~\ref{eso}) that contributes at negative velocities.

\subsection{Funnel or Wavy sheet?}
In the IIIaJ image in Fig.\@~\ref{eso} RCW 37 exhibits a curved
morphology but there is also a parallel line of emission to the
west. This led \citet{redman.et.al00} to suggest that together with
the eastern side, the structure may form a tube/funnel for hot gas
to be vented from the remnant. Just such a structure is present in the
Crab SNR (e.\@g.\@~\citealt{fesen&staker93}) and also in
the more evolved remnant DEM 34a \citep{meaburn87}. The previous
kinematic data were not sufficient to test this possibility. The new
kinematic data across RCW 37 are nearly exactly those expected of a
funnel of circular cross-section undergoing radial expansion. However,
for a funnel, there should be a similar velocity splitting from the
western edge towards the central of the putative funnel as for the
eastern edge and so the ionization of the funnel would have to be
incomplete. Another difficulty with this interpretation is that such a
funnel would be pointing to the centre of the young SNR. Also the
soft RASS X-ray maps reveal a more extensive bow shaped feature and it is
only the hard RASS X-ray component of this that is adjacent to RCW 37
\citep{aschenbach98}.

A plausible alternative interpretation is that the nebula gas is in
the form of a thin `wavy sheet' \citep{hester87} of emission which at
the eastern side overlaps along the line of sight. At the western side
the single thin sheet curves to become edge on, forming the western
edge. Interestingly, whether the single sheet curves into or out of
the plane of sky is ambiguous kinematically. The whole system appears
to be undergoing a bulk expansion generating double-peaked line
profiles at the eastern side where the sheet overlaps itself.

\subsection{Association with \velasnr}

\citet{redman.et.al00} 
showed that RCW 37 and D/D' are spatially coincident but the RASS data
(see Fig.\@~\ref{esorass}) were not of high enough resolution to allow
further investigation and the unlikely possibility of a chance
alignment could not be ruled out. The new Chandra data of D/D'
displayed in Fig.\@~\ref{esochandra} show morphological similarities
to the optical data (see \citealt{plucinsky.et.al02} for an excellent
image). There is an edge to the X-ray emission from D/D'
that runs parallel to the bright optical edge of RCW 37 and the X-ray
peak is close to the peak of the optical emission. A chance alignment
can be ruled out in the light of this new
data. \citet{plucinsky.et.al02} argue that the morphology of D/D'
makes it unlikely that the object is a discrete `bullet' of ejecta and
favour a shock break-out model where a change in density ahead of the
old Vela SNR boundary has led to a localised distortion in the
shock-front.

More controversial is whether, as suggested by \citet{redman.et.al00},
RCW 37 and D/D' are physically associated with \velasnr. They argued
that RCW 37 and D/D' represent a venting of hot gas from the interior
of the remnant to beyond the roughly circular shell as delimited in
the X-ray. This view differs from that of \citet{slane.et.al01b} and
\citet{plucinsky.et.al02} who argue that the X-ray spectrum of D/D'
shows enhanced abundances of O and Ne and is thermal in contrast to
the spectrum of the main remnant body which is non-thermal. This led
\citet{slane.et.al01b} to suggest that RCW 37 and D/D' are not
associated with \velasnr. However, as \citet{plucinsky.et.al02} note,
non-equilibrium ionisation models do not yield such enhanced
abundances. Non-equilibrium models have to be used when the ionisation
timescale is longer than the dynamical timescale as is likely to be
the case for younger remnants. This is clearly important for RCW 37
and D/D' since the dynamical timescale is either 12,000 years or 1000
years depending to which SNR it is
attributing. \citet{plucinsky.et.al02} calculate an ionisation
timescale range of $7.0\times 10^{10} < n_{\rm e}t < 7.0\times
10^{11}~{\rm cm^{-3}~s}$ and find an upper limit of $4.0\times
10^{11}~{\rm cm^{-3}~s}$ by assuming an ambient density of $1~{\rm
cm^{-3}}$ and a timescale of 12'000 years. However, this density is
likely to be an underestimate since \citet{redman.et.al00} used \SII\
line ratios to measure the {\it local} electron density within RCW 37,
finding values of $n_{\rm e}\sim 10^2~{\rm cm^{-3}}$. The ambient
density will have been up to a quarter of this value if there is
little post-shock cooling but less if cooling and compression has
occurred. Clearly, an ambient medium density of $\sim 10~{\rm cm^{-3}}$
and a timescale of $\sim 1000~{\rm yr}$ or less would also be
consistent with the
\citet{plucinsky.et.al02} data. It seems reasonable to suggest therefore 
that the differing spectral properties between the main shell and D/D'
are due to non-equilibrium ionisation effects. A reason for this, that
D/D' has impacted the old wall of the Vela SNR is discussed below.

\citet{combi.et.al99}, using the data of \citet{duncan.et.al96}, 
mapped radio emission from \velasnr\ and have shown that a patch of
non-thermal radio emission from the remnant beyond the X-ray shell
also coincides with D/D' and RCW 37 (see their figure
1). Interestingly, they identify a further three patches of radio
emission beyond the main X-ray remnant boundary. This could indicate
that the X-ray shell is fragmented and interior gas is escaping
through several points. This would strengthen the case that RCW 37 and
the emitting gas giving rise to the X-ray source D/D' are indeed
associated with \velasnr. However, the region is very confused and
\citet{duncan&green00} argue that two of the features that
\citet{combi.et.al99} identify are part of an (unrelated) extended
structure and that the third (feature C, in the \citealt{combi.et.al99}
nomenclature) is isolated and unconnected to
\velasnr. \citet{duncan&green00} suggest that none of the
\citet{combi.et.al99} features are associated with \velasnr. Follow-up
observations will likely clarify this important issue.

\subsection{RCW 37, \velasnr\ and the old Vela SNR} 
The velocity structure and X-ray properties described above may be
explained in the following simple way. We suggest that \velasnr\
exploded within the main Vela SNR. The Vela SNR is approximately
$11,000~{\rm yrs}$ old and by this stage it will have swept up the
interstellar medium into a cold dense shell at its boundary. The
unusual optical morphology, and X-ray properties of RCW 37 and D/D'
are simply due to a localised venting of hot gas from
\velasnr\ impacting the old wall of the Vela SNR. This creates an
X-ray `hot spot' and a curved sheet of optical emission tracing the
inside edge of the old shell wall. \citet{dubner.et.al98} have mapped
the whole Vela SNR in H~{\sc i} and find it to be a located in a thin
shell closely correlated with the X-ray emission of
\citet{aschenbach.et.al95}. This is the wall with which we conjecture the
blast wave is interacting. The hot gas may even have caused localised
expansion of the wall at this point - the eastern boundary of the Vela
SNR as mapped in soft X-rays by \citet{aschenbach.et.al95} is
distorted at the position of RCW 37 and D/D'.

The differing X-ray spectral properties of feature D/D' from the
interior of the \velasnr\ led \citet{slane.et.al01b} to conclude that
the two objects are unrelated to each other (see above).  However, in
the context of the above scenario, differing X-ray properties are
expected since the X-ray gas in fragment D/D' has only recently
impacted the dense neutral wall of the older Vela SNR resulting in
non-equilibrium ionisation. The new Chandra data are indicative of a
shock break-out \citep{plucinsky.et.al02} which in our model would be
caused by the impact of the X-ray emitting gas with the neutral
wall. Interestingly, to the north of the main body of
\velasnr\ there is another much fainter X-ray fragment than D/D'. It
coincides with a knot of non-thermal radio emission (feature C in
\citealt{combi.et.al99}) but there is no obvious optical emission
associated with it. We would expect that the X-ray spectra of this gas
is more akin to that of
\velasnr\ than to D/D' since it is probably not interacting with a
pre-existing obstacle.

\citet{duncan&green00} 
plot radio surface brightness of known galactic supernovae against
diameter and find that unless \velasnr\ is located at around $1~{\rm
kpc}$, it is extremely faint. \citet{duncan&green00} note the unusual
radio properties of \velasnr\ and, as concluded by others, suggest
that the remnant must be expanding into a hot low density medium. If,
as argued above, \velasnr\ is located within Vela SNR then the hot
interior of the older remnant can provide just such an
environment. The radio faintness of \velasnr\ could have important
implications in the investigation of supernova remnants in extreme
environments (such as the radio SNR in the starburst
galaxy M82 by Pedlar and co-workers, see
e.\@g.\@~\citealt{mcdonald.et.al01,pedlar.et.al99,wills.et.al97})

\subsection{The distance to \velasnr}
Estimates of the distance to \velasnr\ are very uncertain, if it is
not assumed to be physically associated with the old Vela
SNR. \citet{duncan&green00} have summarised some of the
difficulties. \citet{aschenbach98} estimate that the lack of
absorption of the X-ray flux gives an upper limit of $1~{\rm
kpc}$. Comparison with the surface brightness of SN1006 gives a lower
limit of $200~{\rm pc}$. \citet{iyudin.et.al98} date the remnant at
$700~{\rm yr}$ giving a distance of $\sim 200~{\rm pc}$ but this is
based on rather uncertain $^{44}\rm Ti$ yields and an estimated shock
velocity. The calcium decay products of $^{44}{\rm Ti}$ were detected
by \citealt{tsunemi.et.al00}, giving an age of between 630 and
970~${\rm yr}$. \citet{burgess&zuber00} re-examined nitrate abundance
data from Antartic ice cores obtained by \citet{rood.et.al79} which
had peaks that seemed to correspond to historical supernovae. They
suggested that an unidentified peak at $\simeq 700$ could have been
caused by \velasnr\ though of course, there is no indication of the
location in the sky of the event. Note however that the claim by
\citet{rood.et.al79} that SN events can be detected in this way has
been questioned in subsequent studies (e.g. \citealt{herron82},
\citealt{risbo.et.al81}). H~{\sc i} studies cannot constrain the
distance to \velasnr\ because of the highly confused nature of this
region which is close to the galactic plane
\citep{duncan&green00}. 

\citet{slane.et.al01a} use the non-thermal spectrum of portions 
of \velasnr\ with a power-law model of the emission to compare derived
H~{\sc i} column densities with other parts of the Vela
SNR. \citet{slane.et.al01a} acknowledge the uncertainties in then
scaling the differences in column density to obtain a distance but
argue that the higher column density of \velasnr\ implies that it is
located somewhere between the back of the Vela SNR and the Vela
molecular ridge at $1-2~{\rm Kpc}$.

The lack of an obvious compact central object within the remnant also
seemed at first to indicate that \velasnr\ must be located at a large
distance \citep{mereghetti01}. However, an X-ray source (but with no
associated optical counterpart) has recently been discovered
\citep{pavlov.et.al01}. It appears that the central object might be rather
similar to that of Cas A - a radio quiet young isolated neutron
star. However, it is worth noting that there is a 64 ms pulsar with
coordinates within the supernova remnant boundary. Due to the
uncertainties involved in the distance of \velasnr\ an association
cannot be ruled out (Redman \& Meaburn, in prep). The various distance
estimates are not consistent with each other and follow-up work is
clearly required to remove the uncertainties inherent in the different
techniques.

Our optical study offers a clear distance constraint if \velasnr\ was
generated within the Vela SNR. \citet{cha.et.al99} have used optical
absorption lines towards a significant sample of OB stars in the
direction of the Vela SNR to constrain the distance as $250\pm30~{\rm pc}$
with a conservative upper limit of $390\pm 100 {\rm pc}$. Clearly, if
indeed \velasnr\ is located within the older Vela SNR then the distance is
constrained to be $\sim 250~{\rm pc}$. We note that if subsequent
studies firmly indicate that \velasnr\ lies well beyond the old Vela
SNR, then our model can be ruled out.

\section{Conclusions}
New \oiii\ line profiles of RCW 37 have been presented that show the
kinematics of this nebula for the first time. A partial velocity
ellipse is discovered in the pv array of line profiles. The kinematics
and morphology could suggest that the structure of RCW 37 is that of a
thin wavy sheet of optical emission that overlaps itself towards the
eastern edge and is undergoing a systematic expansion. The western
edge curves towards the line of sight but does not appear to form a
complete tube or funnel of emission. We compared this feature with
those found towards other SNRs. The evidence that the RCW 37 optical
nebula and associated X-ray feature, D/D' are in fact part of
\velasnr\ has been discussed and we conclude that it is likely
that this is the case. A simple explanation for the origin of the
morphology of RCW 37 is that
\velasnr\ has occured within the older, larger Vela SNR and that a
portion of the supernova ejecta from \velasnr\ has impacted the
pre-existing cold dense wall of the Vela SNR. The thin sheet of
optical emission then traces out the inside edge of this shocked wall
while the X-ray emission marks shock-heated gas. This model predicts
that the distance to \velasnr\ will be similar to that of the main
Vela SNR which has been recently measured to lie at $250\pm
30~{\rm pc}$.

\section*{Acknowledgements}
We thank the staff of the Anglo-Australian observatory for their
excellent assistance during the observations. We have made use of the
{\it Chandra} Data Archive, Smithsonian Astrophysical Observatory,
Cambridge MA, USA; the UKST archive, Royal Observatory Edinburgh,
Scotland UK; and the {\it ROSAT} Data Archive of the Max-Plank
Institut f\"ur extraterrestrische Physik at Garching, Germany. MPR and
DJH are supported by PPARC.

\label{lastpage}
\end{document}